\documentclass[prd,superscriptaddress]{revtex4}
\usepackage{graphicx}
\usepackage{epsf}
\usepackage{epsfig}
\usepackage{amssymb}
\usepackage{amsmath}

\def\1{{\'{\i}}}

\newcommand{\be}{\begin{equation}}
\newcommand{\ee}{\end{equation}}
\newcommand{\bea}{\begin{eqnarray*}}
\newcommand{\eea}{\end{eqnarray*}}

\begin{document}
\title{Reconstructing the interaction rate in holographic models of dark energy}
\author{Anjan A. Sen\footnote{E-mail address: anjan.ctp@jmi.ac.in}}
\affiliation{Center for Theoretical Physics, Jamia Millia Islamia,
New Delhi 110025, India}
\author{Diego Pav\'{o}n\footnote{E-mail address: diego.pavon@uab.es}}
\affiliation{Department of Physics,  Autonomous University of
Barcelona, 08193 Bellaterra (Barcelona), Spain}
%\date{\today}

\begin{abstract}
We reconstruct the interaction rate of the holographic dark energy
model recently proposed by Zimdahl and Pav\'{o}n \cite{wd} in the
redshift interval $0 < z < 1.8$ with observational data from
supernovae type Ia, baryon acoustic oscillations, gas mass
fraction in galaxy clusters, and the growth factor. It shows a
reasonable behavior in the sense that it increases with expansion
from a small or vanishing value in the long past but starts
decreasing at recent times. The later feature suggests that the
equation of state parameter of dark energy does not cross the
phantom divide.
\end{abstract}

\pacs{95.36.+x; 98.80.-k}

\maketitle

\section{Introduction}
Nowadays it is widely agreed that our Universe is currently
accelerating its expansion -see \cite{reviews} for recent reviews.
However, this consensus does not extend to the agent behind this
acceleration and in fact there are many competing candidates. The
simplest one, a tiny cosmological constant, does nicely well at
the pragmatic observational level- but it entails seemingly
unsurmountable problems on the theoretical side. This is why many
researchers are considering other possibilities, particularly some
or other scalar of tachyon field with a strong negative pressure
high enough to drive cosmic acceleration -these go under the
collective name of ``dark energy" fields.

Among the most recent generic proposals there is a very suggestive
one based on the holographic principle. Loosely speaking, the
latter asserts that the entropy of a system is given by the number
of degrees of freedom lying on the surface that bounds it, rather
than in its volume \cite{principle}. The roots of this principle
are to be found in the thermodynamics of black holes
\cite{pedro-jakob}. Nevertheless, as noted by Cohen {\em et al.}
\cite{cohen}, a system may satisfy the holographic principle and,
however, include states for which its Schwarzschild radius is
larger than system size, $L$. This can be avoided by imposing the
constraint that the energy of the system should not exceed that of
black hole of the same size or, equivalently, $\rho \leq 3 \,
c^{2}/(8\pi \, G \, L^{2})$, where $c^2$ is a (non-necessarily
constant) parameter. In the cosmological context $L$ is usually
taken either as the event horizon radius or the Hubble radius.
(For a quick summary of holographic dark energy see section 3 of
Ref. \cite{wd}).

The model of Ref. \cite{wd} is based in two main assumptions,
$(i)$ dark energy complies with the holographic principle with $L$
identified as the radius of the Hubble horizon, $H^{-1}$, hence
$\rho_{x} = 3c^{2} H^{2}/(8 \pi G)$, and $(ii)$ dark energy and
dark matter do not evolve separately but they interact.
Accordingly, the energy conservation equations are
\\
\begin{equation}
\dot{\rho}_{m} + 3H \rho_{m} = Q \, , \qquad \dot{\rho}_{x} + 3H
(1+w) \rho_{x} = - Q \, ,
\label{consv}
\end{equation}
\\
where $w$ is the equation of state parameter of dark energy,
$p_{x}/\rho_{x}$, which is not constrained to be a constant.
Subscripts $m$ and $x$ are for dark matter and dark energy,
respectively.

It should be noted that for spatially flat universes in the
absence of interaction, $Q = 0$, there would be no acceleration
\cite{wd,dwplb}. Besides, $Q$ must be a positive-definite quantity
for the coincidence problem \cite{paul} to be solved (or at least
alleviated) \cite{alleviated}, and the second law of
thermodynamics to be fulfilled \cite{db}. Further, it has been
forcefully argued that the Layzer-Irvine equation \cite{layzer}
when applied to galaxy clusters reveals the existence of the
interaction \cite{elcio}. To the best of our knowledge, the
interaction hypothesis was first introduced, well ahead of the
discovery of late acceleration, by Wetterich \cite{wetterich} to
reduce the theoretical huge value of the cosmological constant,
and  was first used in connection to holography by Horvat
\cite{raul}. As we write, the body of literature on the subject is
steadily growing -see \cite{wd} and references therein. Most
cosmological models implicitly assume that matter and dark energy
couple gravitationally only. However, unless there exists an
underlying symmetry that would set $Q$ to zero (such a symmetry is
still to be discovered) there is no a priori reason to discard the
interaction. Ultimately, observation will tell us whether the
interaction exists.

Following \cite{wd}, we will write the interaction as $Q =
\rho_{x} \Gamma$, where $\Gamma$ is an unknown, semi-positive
definite, function that measures the rate at which energy is
transferred from dark energy to dark matter. Clearly, as long as
the nature of both dark ingredients of the cosmic substratum
remain unknown, $\Gamma$ cannot be derived from first principles;
however, one can resort to observational data (in our case,
supernovae type Ia (SN Ia), baryon acoustic oscillations (BAO),
gas mass fraction in galaxy clusters and the growth factor) to
roughly reconstruct it. The next section focuses on reconstructing
the dimensionless quantity $\Gamma/3 H$.

\section{Reconstruction}
The evolution equation
\\
\begin{equation}
\dot{r} = (1+r) \, \left[  3 H w \frac{r}{1+r} + \Gamma \right]
\label{revol1}
\end{equation}
\\
for the ratio $r \equiv \rho_{m}/\rho_{x}$ between the energy
densities follows from Eqs. (\ref{consv}) and the above
expressions for $\rho_{x}$ and $Q$. With the help of Friedmann
equation $\Omega_{m} + \Omega_{x} + \Omega_{k} = 1$, in terms of
the usual density parameters $\Omega_{i} = 8\pi G
\rho_{i}/(3H^{2}) \;$ ($i = m, x$), and $\Omega_{k} = - k/(a^{2}\,
H^{2})$, where $k$ stands for the spatial curvature index of the
Friedmann-Robertson-Walker metric, we can write
\\
\begin{equation}
\dot{r} = - 2 H\, \frac{\Omega_{k}}{\Omega_{x}} \, q \,
\label{revol2}
\end{equation}
\\
with $q = - \ddot{a}/(a\, H^{2})$, the deceleration parameter.
Here we have assumed $\rho_{x}\propto H^2$ for holographic
dark energy assuming the horizon as the Hubble horizon.

Likewise, starting from the first of Eqs. (\ref{consv}) and using
Friedmann equation, we get for the equation of state parameter the
expression
\\
\begin{equation}
w(z) = (1+r) \left[ \frac{2}{3} \frac{H'}{H}-1 \right]-
\frac{2}{3} \frac{\Omega_{k}}{\Omega_{x}}\left[ 1 -
(1+z)\frac{H'}{H} \right] \, ,
\label{wz1}
\end{equation}
\\
where $z$ denotes the redshift factor and a prime indicates
derivative with respect to this quantity.

We fit the Chevallier-Polarsky-Linder parametrization
\cite{w(z)}, namely,
\\
\begin{equation}
w(z)= w_{0} + w_{1} \frac{z}{1+z} \, ,
\label{wz2}
\end{equation}
\\
where $w_{0}$ is the present value of $w(z)$, and $w_{1}$ a
further constant, to current data from different observational probes
and subsequently use the fitting values for $w_{0}$ and $w_{1}$ to reconstruct
 the dimensionless ratio $\Gamma/3H$.

As for the data, we resort to the various SN Ia observations in
recent times. In particular we use 60 Essence supernovae
\cite{super}, 57 SNLS (Supernova Legacy Survey) and  45 nearby
supernovae. We have also included the new data release of 30 SNe
Ia detected by the Hubble Space Telescope and classified as the
Gold sample by Riess {\it et al.} \cite{super}. The combined data
set can be found in Ref. \cite{davis}. The total number of data
points involved is 192.

Next we add the measurement of the CMB (Cosmic Microwave
Background) acoustic scale  at $z_{BAO} = 0.35$ as observed by the
SDSS (Sloan Digital Sky Survey) for the large scale structure. This
is the Baryon Acoustic Oscillation (BAO) peak) \cite{sdss}.

We also consider the gas mass fraction of galaxy cluster ,
$f_{gas} = M_{gas}/M_{tot}$, inferred from the X-ray observations
\cite{xrs}. This depends on the angular diameter distance $d_{A}$
to the cluster as $f_{gas} = d_{A}^{3/2}$. The number of data
point involved is 26.

Likewise, the 2dF galaxy redshift survey has measured the two
point correlation function at an effective redshift of $z_{s} =
0.15$. This correlation function is affected by systematic
differences between redshift space and real space measurements due
to the peculiar velocities of galaxies. Such distortions are
expressed through the redshift distortion parameter $\beta$.
Correlation function can be used to measure it as $\beta = 0.49
\pm 0.09$ at the effective redshift of z = 0.15 of the 2dF survey.
This result can be combined with linear bias parameter $b= 1.04\pm
0.11$ obtained from the skewness induced in the bispectrum of the
2dFGRS by linear biasing to find the growth factor $g$ at $z =
0.15$, namely $g  = 0.51 \pm 0.11$ \cite{growth}.

\subsection{The spatially flat case}
\begin{figure}[t]
\centerline{\epsfxsize4.3truein\epsffile{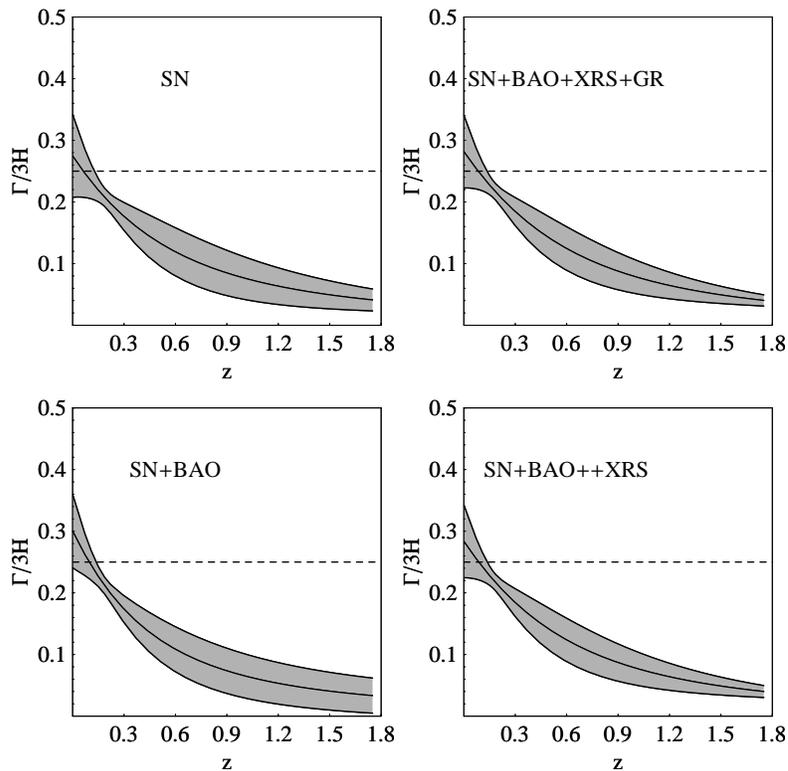}} \caption{The
dimensionless ratio $\Gamma/(3H)$ vs redshift. In the four panels
we have fixed $\Omega_{m0} = 0.25$ and $\Omega_{k} = 0$. The solid
line is for the mean value and the shaded area indicates the
$1\sigma$ region. The region above the horizontal dashed line can
be visited only when the dark energy becomes of phantom type,
i.e., $w < -1$. } \label{fig1}
\end{figure}

\begin{figure}[t]
\centerline{\epsfxsize4.3truein\epsffile{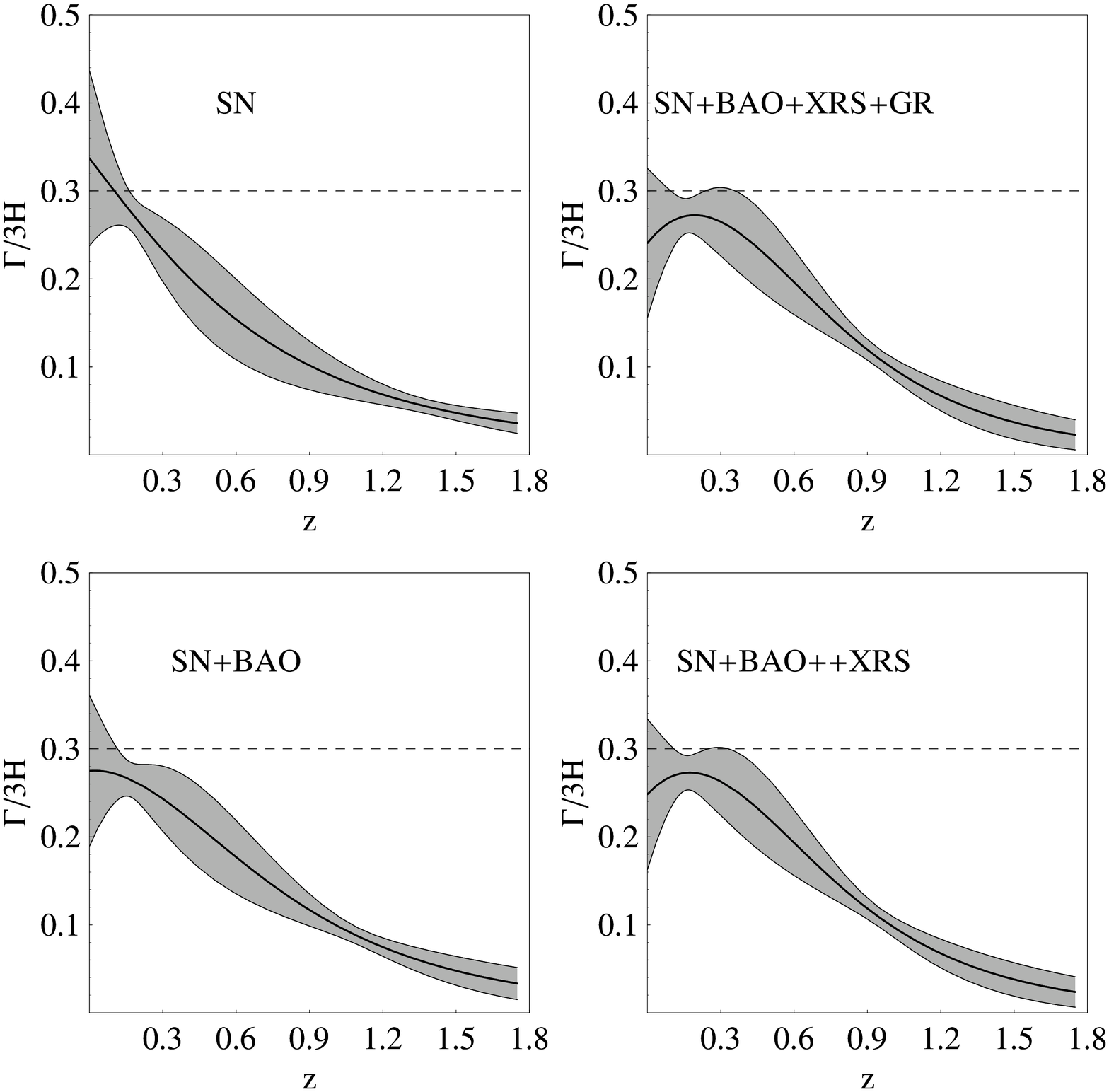}} \caption{Same
as Fig. 1 except that here we have fixed $\Omega_{m0} = 0.30$ and
$\Omega_{k} = 0$.} \label{fig2}
\end{figure}

The simplest case is when $\Omega_{k} = 0$. Then, from
(\ref{revol2}), $r = r_{0}$, where the zero subscript means
present value,  and equations (\ref{revol1}) and (\ref{wz1})
reduce to
\\
\begin{equation}
\frac{\Gamma}{3H} = - r_{0} \left[  \frac{2}{3} (1+z)
\frac{H'}{H}- 1 \right] \, , \label{gammah1}
\end{equation}
\\
and
\\
\begin{equation}
w(z) = (1+ r_{0}) \left[  \frac{2}{3} (1+z) \frac{H'}{H}-
1\right]\, , \label{wz3}
\end{equation}
\\
respectively. Using these two expressions we determine $w_{0}$ and
$w_{1}$ from the data and, with them, we reconstruct $\Gamma/3 H$
-see figures \ref{fig1} and \ref{fig2}.
\\
\\

The best fit values, with $1\sigma$ error bars for the parameters
when all the data (SN Ia + BAO + x-rays + growth factor) are
included, come to be: \noindent $w_{0} = -1.13 \pm 0.24$, $w_{1} =
0.66 \pm 1.35$ (for $\Omega_{m0} = 0.25$ \& $\Omega_{k} = 0$, Fig.
\ref{fig1}); and  $w_{0} = -0.80 \pm 0.28$, $w_{1} = -1.75 \pm
1.79$ (for $\Omega_{m0} = 0.3$ \& $\Omega_{k} = 0$, Fig.
\ref{fig2}).

Here, one cautionary remark seems in order. The fact that $r$ was
never large might lead the reader to think that the model of Ref.
\cite{wd} seriously conflicts with the standard scenario of cosmic
structure formation. One may believe that at early times the
amount of dark matter would have been too short to produce
gravitational potential wells deep enough to lead to the
condensation of galaxies. However, this is not so; a matter
dominated phase is naturally recovered since at high and moderate
redshifts the interaction is even smaller than at present whence
the equation of state of the dark energy becomes close to that of
non-relativistic matter -see \cite{wd} for details.

A related point  is to realize that dark energy clusters similarly
to dark matter when the equation of state of the former stays
close to that of the latter. In this connection, it is worthwhile
to recall the perturbation dynamics of this model. This was
studied in \cite{wd} making use of the perturbed metric
$\mbox{d}s^{2} = - \left(1 + 2 \psi\right)\mbox{d}t^2 +
a^2\,\left(1-2\psi\right)\delta _{\alpha \beta}
\mbox{d}x^\alpha\mbox{d}x^\beta \,$,  with $\psi$ the scalar
metric perturbation, and the Bardeen gauge-invariant variable
\cite{bardeen}
\\
\begin{equation}
\zeta \equiv -\psi + \frac{1}{3}\frac{\hat{\rho}}{\rho + p} =
-\psi - H \frac{\hat{\rho}}{\dot\rho}\ ,
 \label{defzeta}
 \end{equation}
 \\
 which represents curvature perturbations on hypersurfaces of
 constant energy density. Here, an upper-hat means perturbation of
 the corresponding quantity; likewise, $\rho = \rho_{m} + \rho_{x}$ and
 $p = p_{x}$.

Corresponding quantities for the components are
\\
\begin{equation}
\zeta_A \equiv - \psi - H \frac{\hat{\rho}_A}{\dot\rho _A} \quad
\qquad (A = m, x).
\label{defzetaA}
\end{equation}

 On large perturbation scales we have
 that
 \\
 \begin{equation}
\dot\zeta = - H \left(P - \frac{\dot{p}}{\dot{\rho}}D\right)
 \label{dotzetageneral}
\end{equation}
\\
with $P \equiv \hat{p}/(\rho +p)$ and $D \equiv \hat{\rho}/(\rho
+p)$ and parallel expressions for $\zeta_{m}$ and $\zeta_{x}$.
Therefore,  insofar as both equations of state do not differ
significantly the evolution of these two perturbations will be
alike.

In the particular, simplest, case of $\Gamma = {\rm constant}$
last equation reduces to
\\
\begin{equation}
\dot{\zeta} = - \frac{\Gamma}{6(1 + \frac{p}{\rho})}
\frac{\hat{r}}{r^{2}} \,  .
\end{equation}
\\
This one can be readily integrated to (see section 6 of Ref.
\cite{wd} for details)
\\
\begin{equation}
\zeta = \zeta_{i} - \frac{\Gamma}{3}\frac{\hat{r}}{r}\frac{1}{3
H_{i}r - \Gamma}\left[\left(\frac{a}{a_{i}}\right)^{3/2} -
1\right]
 \ .
\label{zeta}
\end{equation}
\\
Again, so long as the the equation of state parameter of dark
energy $w$ remains close to that of dark matter, both components
will cluster in a similar fashion. Further, a non-vanishing
interaction introduces a non-adiabatic feature that grows as
$a^{3/2}$ which will have an impact on the integrated Sachs-Wolfe
effect. Possibly this feature might be used in the future to
discriminate the model under consideration from the $\Lambda$CDM
model -recall that in the latter $\zeta$ remains constant and does
not produce the said effect.

\subsection{Non-spatially flat cases}
When $\Omega_{k} \neq 0$ the ratio $r$ between energy densities is
no longer a constant. This is an extra unknown function in our
fitting procedure. But one should not expect a large variation in
$r$ in the redshift range that has been considered in this paper.
In our subsequent computation, we take the Taylor series expansion
for $r$ around its present day value and take up to the first
order term in the expansion. We therefore parameterize it as
\\
\begin{equation}
r = r_{0} + r_{1} (1-a)= r_{0} + r_{1} \, \frac{z}{1+z} \, , \label{r(z)}
\end{equation}
with  $r_{0}$ is the present day value for $r$. $r_{1}$ is a constant
which can be related to the present ratio
of densities between $\Omega_{k}$ and $\Omega_{x}$ by
\\
\begin{equation}
\frac{\Omega_{k0}}{\Omega_{x0}} = - \frac{r_{1}}{2} \left[1 -
\left(\frac{H'}{H}\right)_{z = 0} \right]\, .
\label{presentdensitiesratio}
\end{equation}

This can be used to fix the unknown constant $r_{1}$ for a given
$\Omega_{k0}$ and $\Omega_{x0}$.  Also
\\
\begin{equation}
\frac{\Gamma}{3H} = - \frac{1}{1+r} \, \left(r' \, \frac{1+z}{3}-
w\, r \right) \, \label{gammah2}
\end{equation}
\\
where $w$ is given by Eq. (\ref{wz1}).

\begin{figure}[t]
\centerline{\epsfxsize4.3truein\epsffile{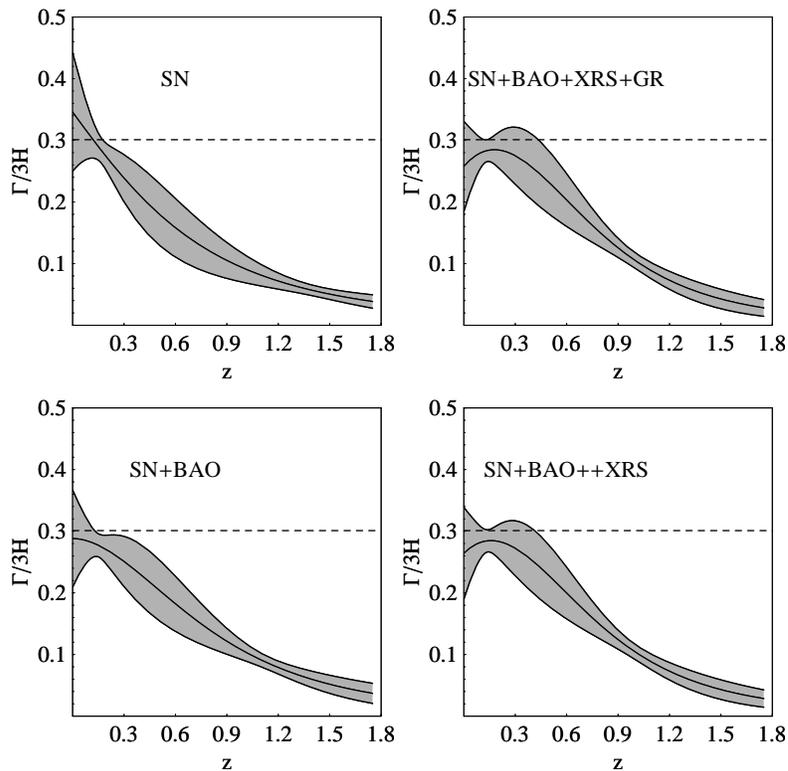}} \caption{Same
as Fig. 1 except that we have fixed $\Omega_{m0} = 0.30$ and
$\Omega_{k0} = 0.002$.} \label{fig3}
\end{figure}

Using these expressions, the ratio $\Gamma/3H$ is reconstructed
from the data in Fig. \ref{fig3}.
\\

The best fit values, with $1\sigma$ error bars for the parameters
when all the data (SN Ia + BAO + x-rays + growth factor) are
included, come to be: $w_{0} = -0.806 \pm 0.29$, $w_{1} = -1.74
\pm 3.33$. It is seen that the curvature, being small as WMAP 3yr
\cite{wmap3} tells us, has little consequence on the evolution of
the interaction rate.

\section{Concluding remarks}
We reconstructed the interaction term $Q$ of Ref. \cite{wd} in the
redshift interval  ($0< z< 1.8$) of observational data (supernovae
type Ia, baryon acoustic oscillations, gas mass fraction, and
growth factor). The interaction rate $\Gamma$ (and hence $Q$) is
always positive, its general trend is to decrease as $z$ increases
but it shows no indication of becoming negative at larger
redshifts. This corroborates that as previously suggested
\cite{db,elcio} the energy transfer proceeds from dark energy to
dark matter rather than otherwise. While phantom behavior cannot
be excluded at recent and present times it only occurs in a
manifest way either for large $\Omega_{x0}$ -see Fig. \ref{fig1}-
or when just the supernovae data are considered (top-left panel of
Figs. \ref{fig1}-\ref{fig3}). When $\Omega_{x0}$ is a bit lower
(say, $0.7$) and BAO and other data are included, the mean value
of dimensionless  interaction rate,  $\Gamma/3H$, no longer
crosses the phantom divide (i.e., the horizontal dashed line). It
simply reaches a maximum near $z = 0$ and decreases with
expansion. This is a reassuring result as holography is not
compatible with phantom energy \cite{bak}. On the other hand, it
should be noted that $\Omega_{x0}$ values as high as $0.75$ do not
seem favored from a combination of results from WMAP 1yr and weak
lensing which yields $\Omega_{x0} = 0.70 \pm 0.3$ \cite{contaldi}.

Adding a small curvature term -say, $\Omega_{k0} = 0.002$-, has
only a tiny  impact (compare Figs. \ref{fig2} and \ref{fig3}).
This also holds true when the curvature bears the opposite sign;
this is why we have not included a corresponding figure.

In any case, it should be noted that the concordance $\Lambda$CDM
model ($w_{0} = -1$, $w_{1} = 0$) shows compatibility within
$1\sigma$ confidence level with the set of data considered in this
work.

\acknowledgments{This research was partially supported by the
Spanish Ministry of Education and Science under Grant
FIS2006-12296-C02-01, and the ``Direcci\'{o} General de Recerca de
Catalunya" under Grant 2005 SGR 00087. A.S. acknowledges financial
support from the ``Universitat Au\`{o}noma de Barcelona" through a
grant UAB-CIRIT, VIS-2007, for visiting professors.}

%%%%%%%%%%%%%%%%%%%%%%%%%%%%%%%%%%%%%%%%%%%%%%%%%%%%%%%%%%%%%%%%%%%%%%%%%%%%%%%

\end{document}